Schulz, Thimo; Speck, Christina (2026). Conceptualising Reflective Use: Toward a Process Perspective on Human-AI Interaction.
*Proceedings of the 34th European Conference on Information Systems (ECIS 2026).*

# CONCEPTUALISING REFLECTIVE USE: TOWARD A PROCESS PERSPECTIVE ON HUMAN-AI INTERACTION

*Short Paper*

Thimo Schulz, Karlsruhe Institute of Technology, Karlsruhe, Germany, thimo.schulz@kit.edu

Christina Speck, Karlsruhe Institute of Technology, Karlsruhe, Germany, christina.speck@kit.edu

## Abstract

*The rapid diffusion of generative artificial intelligence (genAI) systems reshapes how individuals engage with information systems, requiring users to monitor, assess, and adapt their interaction with non-deterministic systems. Existing constructs capture elements of this engagement but do not account for the situated dynamics of the entire evaluative process in genAI use. This research-in-progress, situated in a larger endeavour towards a scale development, derives an initial conceptualisation of reflective use: a behavioural-knowledge capability that unfolds across pre-use, in-use, and post-use phases, reinforced through situated reflective knowledge gained in practice. Drawing on expert interviews and a focus group, we identify four core components of reflective use and show how they form an iterative capability cycle anchored within the motivational needs outlined in self-determination theory. Understanding reflective use is essential to ensure appropriate reliance and high decision quality, and thus provides a foundation for promoting responsible and effective human-AI interaction.*

## 1 Introduction

The rapid diffusion of generative artificial intelligence (genAI) systems is reshaping how individuals and organisations interact with information systems (IS). Unlike traditional IS that typically process, retrieve, or transform predefined inputs, generative systems create novel outputs based on probabilistic models and contextual inference (Feuerriegel et al., 2024; Riemer & Peter, 2024). Their behaviour is characterised by non-determinism and generative novelty, enabling users to accomplish a wide variety of tasks in the same interactive genAI interface, currently often chat-based, and with the same underlying model (Storey et al., 2025). As these models, system capabilities, and interaction patterns continue to evolve over time, users cannot assume stable system behaviour even within the same interface. These characteristics fundamentally change what "use" entails: rather than merely executing predictable commands, users must continuously evaluate whether genAI is appropriate for a task, how it needs to be instructed, how outputs should be interpreted, and to what extent they can be relied upon, while iteratively steering, assessing, and refining system outputs under uncertainty (Banh & Strobel, 2023; Storey et al., 2025). In such interaction settings, genAI systems shift substantial responsibility to users, who must manage their engagement dynamically to prevent over/under-reliance, misuse, or incorrect outcomes. Consequently, effective interaction with genAI systems requires not only technical proficiency, but also a form of reflective engagement through which users regulate and adapt whether and how they use the system across the interaction process.

Reflection-in- and -on-action as foundational perspectives (Schön, 2017) are particularly useful here because they foreground reflection as something that can accompany, shape, and revise action rather than merely follow it. In this tradition, reflection is not confined to retrospective evaluation after action, but also encompasses anticipatory and concurrent forms of evaluative engagement through which individuals consider how to act, monitor unfolding action, and revise their judgments in practice. Related constructs, e.g., mindful IT use (Thatcher et al., 2018), appropriate reliance (Eckhardt et al., 2026), or effective use (Burton-Jones & Grange, 2013), capture specific aspects of reflection and use, but do not account for the situated, processual dynamics of genAI use (cf. Table 1). These concepts provide valuable foundations, yet none offers a contextualised understanding of why and how users assess and adapt their engagement with genAI, which is essential for ensuring appropriate reliance and high decision quality. Yet IS research lacks a construct that captures the behavioural ways in which individuals regulate their genAI use, as well as the underlying capability that enables such behaviour. We therefore pose the following research question (**RQ**):

*How can reflective use of generative AI systems be conceptualised, and which behaviours, processes, and motivational elements constitute this construct?*

To address this gap, we develop a theoretically grounded conceptualisation of reflective use, in a larger endeavour toward constructing a scale to assess reflective genAI use (MacKenzie et al., 2011). In this research-in-progress paper, we draw on expert interviews and a focus group and follow the Gioia (2013) methodology and grounded-theory principles (Glaser & Strauss, 2017) to inductively derive the foundational dimensions of reflective use. In this process, motivational dynamics emerged as central, prompting us to integrate self-determination theory (SDT) as a complementary theoretical lens (Ryan & Deci, 2000). This paper presents the resulting conceptualisation of reflective use in genAI interaction.

## 2 Conceptual Neighbourhood and Theoretical Background

A central step in conceptual development is to clearly demarcate the focal construct from related concepts (MacKenzie et al., 2011). GenAI use interacts with multiple literatures: IS use, cognitive psychology, education, and reflective practice. To provide conceptual clarity, we offer an overview of neighbouring constructs and how they differ from reflective use (cf. Table 1). As this is a research-in-progress, the overview is intentionally selective and centres on the constructs most critical for distinguishing reflective use at this stage of development.

Several constructs describe forms of thinking practices but do not account for the dynamics of genAI interaction. **Reflection-in-action** and **reflection-on-action** (Schön, 2017) capture how individuals adjust their thinking during and after practice, but they remain broad, practice-oriented descriptions rather than contextualised accounts of reflection in interaction with opaque, probabilistic systems. **Critical thinking**, **reflective thinking**, and **metacognition** (Flavell, 1979; Lai, 2011; Rodgers, 2002) describe internal reasoning, monitoring, and evaluation, yet they are not tool-mediated. These constructs therefore inform the cognitive logic of reflection but do not specify how reflective behaviour unfolds.

Other related concepts describe relatively stable orientations toward technology. **Mindful IT use** (Thatcher et al., 2018) captures a dispositional orientation toward alert, flexible IT engagement, yet it is a trait, not a behavioural process unfolding across interaction phases, and does not incorporate reflective knowledge development. **AI literacy** (Pinski & Benlian, 2024) likewise represents a knowledge-based competence (what users know in principle) rather than how they adapt their behaviour applying it. While such traits may enable reflective use, they do not capture the behavioural regulation intrinsic to it.

A final set of constructs focuses on the results of system engagement rather than on the reflective processes themselves. **Appropriate reliance** (Eckhardt et al., 2026) focuses on whether users follow correct AI advice and disregard incorrect advice, offering a behavioural calibration perspective. However, it remains centred on discrete advice-taking decisions and does not model the ongoing reflection and knowledge evolution that characterise reflective use of genAI systems. Thus, while related, appropriate reliance captures only one specific behavioural outcome rather than a broader reflective capability. **Effective use** (Burton-Jones & Grange, 2013) explains the extent to which users apply a system in a way that appropriately leverages its features to achieve desired task outcomes. It is an outcome-oriented construct concerned with whether system use produces high-quality performance, rather than with the reflective behaviours through which users evaluate, steer, or learn from interaction. It thus represents a result of reflective engagement rather than its underlying behavioural processes.

| | Construct | Description | Differentiating Feature | Example |
|---|---|---|---|---|
| | Reflective Use | A behavioural-knowledge capability involving deliberate pre-use, in-use, and post-use reflection | Situated in human-AI interaction; integrates behavioural phases, evolving knowledge, and motivational reinforcement; focussed on genAI | This study |
| Thinking Practices | Reflection in & on Action | Reflective thinking during and after practice | Broad, practice-oriented descriptions; not formalised or contextualised for socio-technical, opaque, probabilistic systems | (Schön, 2017) |
| | Critical / Reflective Thinking | Deliberate, self-regulated judgment and evaluation | Focuses on reasoning and argumentation; not tool-mediated; does not address calibration of system outputs | (Lai, 2011; Rodgers, 2002) |
| | Meta-Cognition | Awareness, monitoring, and regulation of one's cognition | Concerned with internal self-regulation; lacks socio-technical embedding and does not address interaction with system behaviour | (Flavell, 1979) |
| Trait-likes | Mindful IT Use | A secondary, dynamic IT-specific trait describing alert, flexible interaction styles | Stable orientation; trait-based and not modelling reflection as a behavioural process or knowledge evolution | (Thatcher et al., 2018) |
| | AI Literacy | Knowledge, competencies, and attitudes for understanding, using, and evaluating AI | Trait-like and concept-oriented; does not describe situated, real-time behavioural regulation or reflection during interaction | (Pinski & Benlian, 2024) |
| Outcome-focused | Appropriate Reliance | Behaviour of accepting correct AI advice and rejecting incorrect advice | Focused on discrete advice-taking in decision tasks; does not model ongoing reflection, calibration, or knowledge evolution | (Eckhardt et al., 2026) |
| | Effective Use | Using a system in a way that fully leverages its features for task accomplishment | Outcome-focused; does not explain the reflective processes through which users evaluate, steer, or learn from non-deterministic systems | (Burton-Jones & Grange, 2013) |

*Table 1. Distinguishing reflective use from related constructs.*

Across these literatures, we find no construct that integrates reflective behaviour, situated knowledge, and motivational reinforcement within the socio-technical and non-deterministic dynamics of genAI. What is missing is a construct that integrates these perspectives into a coherent and holistic understanding of reflective behaviour for genAI use. This gap motivates our development of reflective use as a contextualised extension of reflective practice tailored to human-AI interaction, often mediated through chat-based, textual, voice, or multimodal interfaces.

**Self-determination theory** provides an additional theoretical background that complements the construct boundaries outlined above. While not a neighbouring construct, SDT helps illuminate the motivational mechanisms that underpin reflective engagement. SDT posits that autonomy, competence, and relatedness facilitate self-regulated action (Ryan & Deci, 2000). Prior IS research has used this perspective to explain why users engage with technologies when these support discretion in use, a sense of effectiveness in task accomplishment, and socially embedded forms of interaction and learning (Karahanna et al., 2018; Lee et al., 2015; Rezvani et al., 2017). Emerging genAI-related work similarly suggests that genAI can shape users' perceived autonomy by extending their room for action, competence by strengthening their sense of capability in knowledge work, and relatedness by affecting how they experience support and connection in their work practices (Oesinghaus et al., 2024). At the same time, this literature also indicates that such effects are not unconditional: the same technologies may reduce autonomy when they pre-structure choices, weaken competence when they are experienced as substituting rather than extending one's own capabilities, or alter relatedness by displacing or reshaping interaction with others. Reflective use becomes relevant precisely in this tension, because the extent to which genAI supports rather than undermines these psychological needs depends on whether users deliberately determine when and how to employ it, critically monitor and adjust their interaction in use, and reflect on its consequences for future action. SDT thus offers a theoretical anchor for understanding why reflective use may develop and be sustained as an iterative capability, but it does not replace the need for a dedicated construct that captures the behavioural and knowledge-oriented processes of genAI interaction.

# 3 Methodology

To conceptualise reflective use of genAI, we adopted a qualitative and inductive research design suitable for generating nascent theoretical insights in emerging IS phenomena. Our approach follows grounded-theory principles as articulated by Glaser & Strauss (2017) and the systematic inductive procedure proposed in the Gioia methodology (Gioia et al., 2013). We relied on expert interviews and a subsequent focus group workshop to iteratively surface, refine, and validate the emerging conceptual structure.

## 3.1 Data Collection

We conducted three semi-structured interviews with experts experienced in genAI use and integration. Participants were recruited via purposive sampling to ensure theoretical relevance (Patton, 2014) and included a Chief Digital Officer (I1), an academic specializing in AI-related competencies (I2), and a professor of human-centric AI (I3). None were affiliated with the author team. Interviews lasted 30-50 minutes, were conducted online, recorded with consent, and transcribed verbatim. The semi-structured protocol was based on the conceptual landscape and explored (a) where reflective use begins and ends, (b) behaviours characterizing reflection, and (c) indicators of this behaviour. Questions and probes consistently centred on interactive genAI situations in which users engage directly with genAI during task accomplishment and remain involved in judging appropriateness, interpreting outputs, and adapting their course of action. Semi-structured interviews allowed comparability across interviews and flexibility for deeper elaboration (Adams, 2015).

To refine and critically assess the initial conceptualisation derived from the interviews, we conducted a focus group discussion with four additional experts, independent of the author team (Morgan, 1997). Two senior academics working on human-centred systems and two industry managers responsible for implementing genAI applications provided complementary perspectives. The focus group took place as part of a larger three-hour workshop on responsible and effective genAI use. The reflective-use segment lasted approximately 60 minutes. We presented a preliminary conceptualisation and introduced its components before inviting open discussion. Similarly to the interviews, the discussion centred on interactive genAI use situations in which users engaged in interactions and remained actively involved in deciding whether and how to use the system, interpreting outputs, and revising subsequent action. Following established guidance for qualitative focus groups, the session was guided but not strictly structured to encourage critical reflection and elaboration (Morgan, 1997). The session was not audio-recorded; extensive notes were taken by the authors.

## 3.2 Data Analysis and Conceptualisation

Interview data analysis followed the Gioia methodology (Gioia et al., 2013), which offers a rigorous, inductive approach to articulating new constructs. In the first-order analysis, we stayed close to participants' terms before abstracting them into second-order themes and finally aggregating dimensions that captured the emerging structure of reflective use. Two authors of this paper conducted the coding, using MaxQDA to organize the data, and supported by a third researcher who contributed to discussions and sense-making. Following consensus-oriented coding practices (Kuckartz, 2014), we iteratively compared interpretations, discussed coding decisions, and refined emerging themes through repeated analytic cycles. This approach reflects grounded-theory principles of constant comparison (Glaser & Strauss, 2017) while allowing flexibility to capture the evolving conceptual landscape.

Data from the focus group were analysed using an interpretive, synthesis-oriented approach consistent with established guidance for qualitative focus-group research (Morgan, 1997). As the session was not recorded, we relied on extensive contemporaneous notes documenting key statements, points of agreement or divergence, and collective meaning-making processes. These notes were reviewed collaboratively and compared against the preliminary conceptualisation developed from the interview data. The analysis focused on identifying refinements, clarifications, and boundary conditions that participants collectively articulated. This interpretive review informed the subsequent revision of the conceptual structure and served to validate and extend the interview-based insights.

# 4 Results

Our analysis produced a multi-layered conceptualisation of reflective use grounded in over 150 coded interview passages. Following the Gioia (2013) methodology, we present the inductive structure derived from the interviews, then how the focus group refined and theoretically anchored the emerging construct.

## 4.1 Interview-Based Findings: Dimensions of Reflective Use

The Gioia analysis surfaced four aggregate dimensions of reflective use. Three relate to behavioural phases (pre-use, in-use, and post-use reflection) while a fourth represents knowledge that both enables and results from these behaviours. Two additional sets of factors (affective/dispositional and contextual conditions) influence but do not constitute reflective use. Figure 1 summarises the data structure.

**Pre-use reflection: Intentionality and awareness.** Participants consistently emphasised that reflection begins before interacting with the system. Users consider whether and why genAI should be used, paying attention to task sensitivity, domain constraints, and stakes: "Reflective use begins at the point where I'm questioning if a usage makes sense in this context" (I1). Clarifying purpose and expectations was seen as foundational for later evaluation: "Reflection requires setting a purpose or expectation beforehand" (I2); "What is my intention for using this tool?" (I3).
**In-use reflection: Cognitive and metacognitive engagement.** During interaction, reflective users critically and consciously engage with AI-generated content. This involves evaluative behaviours such as "engaging critically with the result, but on a content level rather than merely on a process level" (I2) and deliberately slowing down automatic responses: "Many tasks run subconsciously and quickly, but here you need Type 2 thinking: slowing down to reflect" (I3). Participants also emphasised metacognitive monitoring – "questioning whether one's judgment is correct and how the AI may have influenced it" (I1) – as well as recognising how genAI can reshape one's reasoning and opportunities as "it enables achieving higher goals and does not take something away. This is metacognition." (I2). Finally, reflective users calibrate their reliance on outputs, with experts noting persistent "tendencies toward over- or under-reliance" (I3).

**Post-use reflection: Adaptive learning and judgment revision.** Reflection extends beyond the immediate interaction into a learning-oriented phase that involves reviewing outcomes and adjusting strategies. Experts described evaluating what worked, what did not, and how to change future approaches: "Where was the benefit or the problem? What do I have to change next time?" (I2). These activities capture how users consolidate experience over time, thereby increasing effectiveness.

**Enabling and emerging from behaviour: Reflective knowledge.** Beyond behaviours, experts stressed knowledge about model behaviours, limitations, and appropriate use contexts. While AI literacy is "necessary but not sufficient" (I1), users acquire situated knowledge through practice, allowing them to "judge based on experience where genAI is a suitable tool" (I2) and draw on social knowledge, e.g., knowing whom to consult, as "it is important to be able to discuss and reflect in a team" (I2).

**Facilitator & Inhibitors.** Interviews also surfaced factors that facilitate or hinder reflective use. At the individual level, traits such as curiosity, openness, and confidence support reflective engagement, whereas anxiety, time pressure, or low affinity toward AI inhibit it. At the contextual level, norms, shared standards, and leadership practices within organizations shape opportunities for reflection. While these factors strongly influence reflective use, they were not described as constitutive of the construct itself. Instead, they represent contextual preconditions that enable or constrain reflective behaviour.

## 4.2 Focus Group Refinement and Theoretical Anchoring

The focus group validated the interview-based structure and offered three key refinements: (1) clarifying the relationship between behaviour and knowledge, (2) describing reflective use as an iterative capability development process, and (3) aligning the behavioural dimensions and phases with motivational dynamics from self-determination theory.

Focus group participants agreed that reflective use comprises both deliberate reflective behaviour and the knowledge that forms through repeated reflective engagement. While AI literacy provides foundational awareness, participants argued that reflective knowledge is more situated and experiential and develops through practice rather than instruction. This distinction helped further separate reflective use from neighbouring constructs such as literacy or simple critical thinking. Participants viewed affective and contextual influences as important enablers, but not as components of the construct itself. This helped delineate the construct's core: a behavioural process that builds and relies on evolving knowledge. Participants emphasised that reflective use matures through a feedback loop. Pre-use, in-use, and post-use reflections collectively generate knowledge about when and how genAI can be used effectively. This knowledge, in turn, shapes future reflective behaviour, making it more intuitive, efficient, and context-sensitive. This iterative cycle is a core extension of the model beyond the interview-based Gioia structure and captures the dynamic nature of reflective use. Another central contribution of the focus group was the further interpretation of reflective use as not only a behavioural-knowledge capability, but also as one shaped by motivational dynamics. In discussing how reflective use develops over time, participants repeatedly pointed to users' discretion in deciding whether and how to employ genAI, the gradual development of confidence and capability through iterative engagement, and the importance of exchange with others when comparing experiences and revising use practices. These patterns suggested a close connection to the psychological needs described in SDT (Oesinghaus et al., 2024; Ryan & Deci, 2000) and offer a complementary theoretical lens for understanding how reflective use may be reinforced over time. Integrating this perspective, we derive a first conceptualisation presented in section 4.3.

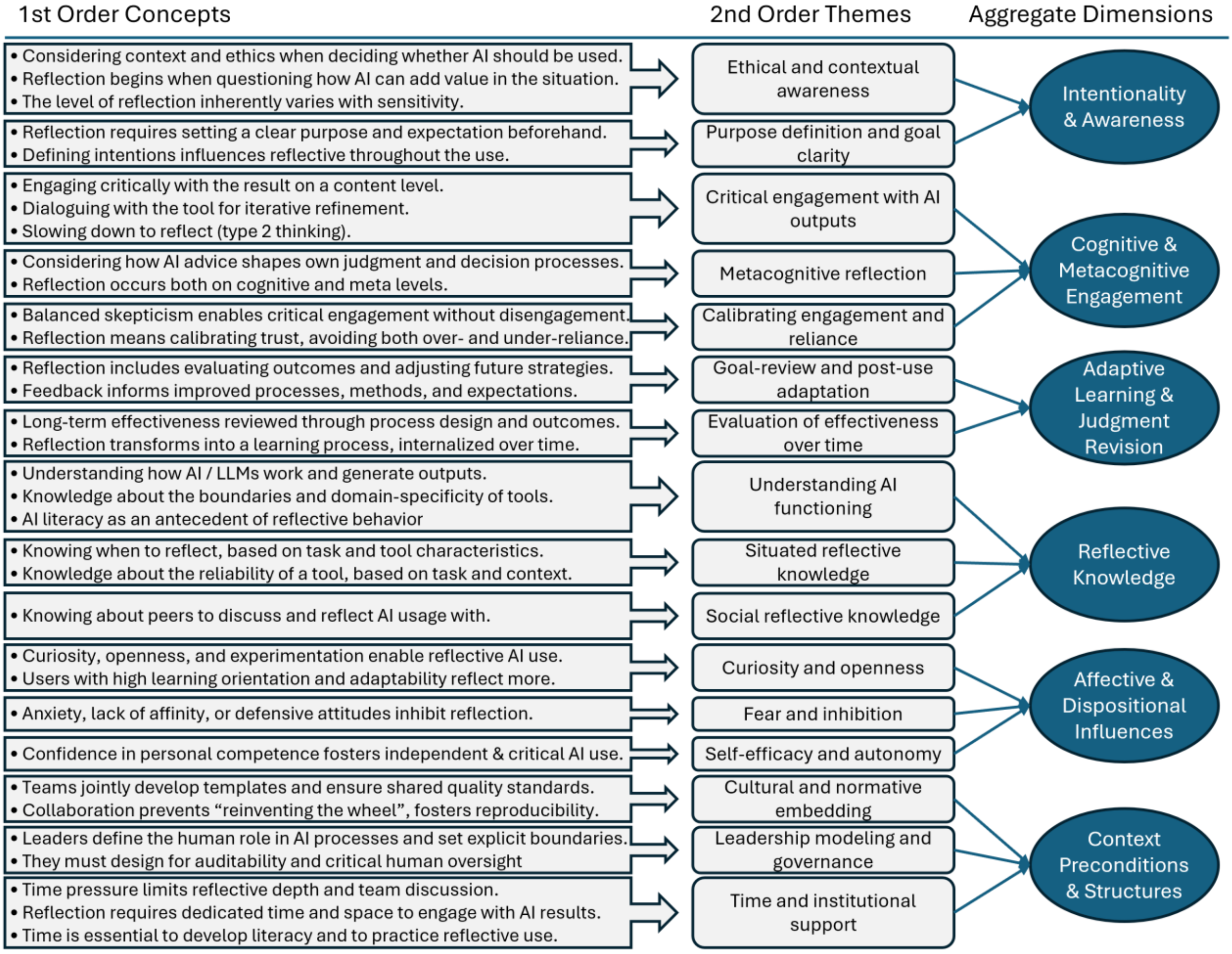

*Figure 1. Gioia Mapping: Concepts, themes, and dimensions shaping reflective use.*

### 4.3 Conceptualisation of Reflective Use

Integrating insights from both data sources, we conceptualise reflective use as a behavioural-knowledge capability comprising: (1) Intentionality and awareness (pre-use reflection); (2) cognitive and metacognitive engagement (in-use reflection); (3) adaptive learning and judgment revision (post-use reflection); (4) reflective knowledge, which both enables and emerges from these behaviours. While reflective knowledge appears as a distinct component of the construct, its role is inherently dual. First, it functions as an antecedent: Foundational understanding of genAI capabilities, limitations, and risks enables users to initiate and enact reflective behaviour during interaction. Second, reflective knowledge is also an outcome of reflective use, because users acquire situated, experience-based insights through repeated reflection. Over time, this knowledge becomes internalised and shapes subsequent engagement. Reflective knowledge is therefore not an external prerequisite but an integral, recursive part of the capability itself – it both enables reflective behaviour and is continuously shaped by it.

Together, these components form a dynamic cycle (cf. Figure 2) in which reflective behaviour and knowledge continually reinforce one another. SDT helps interpret the motivational dynamics that may support this cycle. Pre-use reflection foregrounds **autonomy** insofar as users determine whether and how genAI should be employed in line with their goals and values. In-use reflection strongly engages **competence**, as individuals monitor, evaluate, and adjust their interaction while seeking effective control over system behaviour. Post-use reflection often connects to **relatedness**, when users compare experiences, discuss outcomes, and learn with or from others. These needs are not confined to single phases but may overlap across the interaction process. Importantly, this alignment surfaced inductively rather than being imposed a priori. SDT therefore provides a theoretical anchoring for understanding how reflective use may be reinforced over time: By supporting autonomy, competence, and relatedness as fundamental psychological needs, reflective use becomes more likely to be sustained and to develop into a self-reinforcing pattern of responsible and effective interaction with genAI. As reflective use is conceived as a behavioural-knowledge capability, the subsequent scale development will operationalise it as a reflective measurement model with observable behaviours serving as manifestations of the underlying construct.

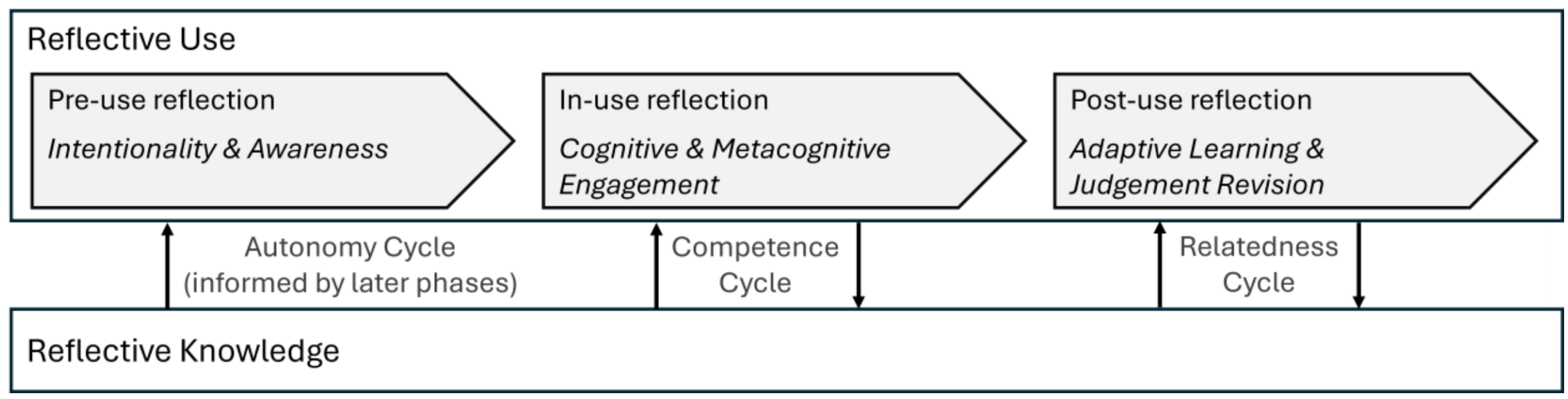

*Figure 2. Conceptualisation of reflective use.*

## 5 Discussion & Conclusion

Answering the posed RQ, this research proposes an initial conceptualisation of reflective use as a behavioural-knowledge capability that captures how individuals regulate their engagement with interactive genAI systems. The construct addresses the shift of evaluative responsibility to users imposed by genAI's non-deterministic, partially opaque, and continually evolving behaviour, thereby capturing how users navigate the dynamic, knowledge-building, and motivationally reinforced process of interacting with generative systems. Reflective use brings these elements together into a single coherent capability tailored to socio-technical characteristics of genAI.

By extending Schön's (2017) reflection-in- and -on-action into the socio-technical setting of genAI, our findings show that reflection unfolds across pre-use, in-use, and post-use phases and is complemented by evolving situated knowledge shaped through practice. This moves beyond trait-like orientations such as mindful IT use (Thatcher et al., 2018) and beyond outcome-oriented constructs such as effective use (Burton-Jones & Grange, 2013) or appropriate reliance (Eckhardt et al., 2026), emphasising instead the process of regulating interaction with a multifaceted, unpredictable tool. Reflective use emerges as an iterative capability, where reflective behaviours generate situated knowledge, which in turn informs subsequent engagement and enables more intuitive, context-sensitive use over time. The alignment between these phases and the motivational needs of SDT (Ryan & Deci, 2000) offers a further theoretical explanation for how reflective use develops and why it may be sustained over time: When reflective engagement supports autonomy, competence, and relatedness, it is more likely to endure and promote more effective and responsible human-AI interaction. While recent genAI-related work has begun to examine how genAI may affect users' autonomy, competence, and relatedness (e.g., Oesinghaus et al., 2024), our contribution shifts attention to the behavioural process through which such motivational potential may be realised in use. In this sense, the contribution of reflective use lies not only in introducing a new label, but in integrating phase-spanning reflective behaviour, situated reflective knowledge, and motivational reinforcement into a single processual capability for genAI interaction.

For practice, the conceptualisation underscores that reflective use requires support before, during, and after genAI engagement. Before use, organisations can support reflective use by providing guidance on when genAI is appropriate, which risks and sensitivities should be considered, and what role the system should play in specific tasks. During use, support may involve prompting and verification guidelines, mechanisms that encourage critical review of outputs, and norms that help users calibrate reliance rather than accept or reject outputs uncritically. After use, organisations can foster reflective learning through debriefs, peer exchange, and opportunities to document and share lessons learned across teams. The reflective knowledge dimension, in turn, suggests that such capability cannot be built through isolated training sessions alone but must develop through continued, supported use.

As a research-in-progress study, our empirical foundation is intentionally modest, comprising three expert interviews and a focus group for early-stage conceptual emergence. This design favours conceptual refinement over empirical breadth. The interview base is therefore limited in size and reflects expert rather than end-user perspectives, while the focus group primarily served to refine and theoretically anchor the emerging conceptualisation. In addition, although the focus group generated rich discussion, the session was not audio-recorded and the analysis therefore relied on extensive contemporaneous notes rather than a full transcript. While additional user perspectives may enrich the nuance of individual dimensions and broaden the empirical grounding, the core structure of the construct is robust across data sources. Prior to ECIS 2026, we will extend our data collection to further refine the conceptual boundaries and support and inform the next steps toward scale development. This includes elaborating the preliminary nomological network of reflective use and developing an initial item pool following MacKenzie et al. (2011).

In sum, this short paper offers a first theoretically grounded conceptualisation of reflective use tailored to the characteristics of genAI. It lays the foundation for operationalising the construct and contributes to emerging IS scholarship seeking to understand how users can meaningfully and responsibly engage with generative systems.